\documentclass[twocolumn,showpacs,aps,prx]{revtex4}

\newcommand{\beq}{\begin{eqnarray}}
\newcommand{\eeq}{\end{eqnarray}}
\newcommand{\beqq}{\begin{eqnarray*}}
\newcommand{\eeqq}{\end{eqnarray*}}

\newcommand{\be}{\begin{equation}}
\newcommand{\ee}{\end{equation}}
\newcommand{\bea}{\begin{eqnarray}}
\newcommand{\eea}{\end{eqnarray}}

\usepackage[normalem]{ulem}
\usepackage{graphicx}
\usepackage{dcolumn}
\usepackage{bm}
\usepackage{color}
\usepackage{tabularx}

\usepackage{amsmath}

\begin{document}

\begin{titlepage}

\title{Electron Teleportation in Multi-Terminal Majorana Islands: \\
Statistical Transmutation and Fractional Quantum Conductance}
\author{Karen Michaeli$^1$,  L. Aviad Landau,$^2$ Eran Sela,$^2$ Liang Fu$^3$}
\affiliation{$^1$ Department of Condensed Matter Physics, The Weizmann Institute of Science,
Rehovot 76100, Israel \\ $^2$ Raymond and Beverly Sackler School of Physics and Astronomy, Tel-Aviv University, Tel Aviv 69978, Israel \\ $^3$ Department of Physics, Massachusetts Institute of Technology,
Cambridge, MA 02139, USA}

\begin{abstract}
We study a topological superconductor island with spatially separated Majorana modes coupled to multiple normal metal leads by single electron tunneling in the Coulomb blockade regime. We show that low-temperature transport in such Majorana island is carried by an emergent charge-$e$ boson composed of a Majorana mode and an electron from the leads. This transmutation from Fermi to Bose statistics has remarkable consequences. For noninteracting leads, the system flows to a non-Fermi liquid fixed point, which is stable against tunnel couplings anisotropy or detuning away from the charge-degeneracy point. As a result, the system exhibits a universal conductance at zero temperature, which is a fraction of the conductance quantum, and low-temperature corrections with a universal power-law exponent. In addition, we consider Majorana islands connected to interacting one-dimensional leads, and find different stable fixed points near and far from the charge-degeneracy point.
\end{abstract}


\maketitle

\draft

\vspace{2mm}

\end{titlepage}

Majorana modes are an unusual type of quasiparticles in topological superconductors, consisting of localized electron and hole excitations in an equal superposition \cite{kitaev,read,wilczek}. The presence of spatially separated Majorana modes in a macroscopic topological superconductor gives rise to degenerate ground states that are locally indistinguishable and topologically protected. In a  mesoscopic superconductor island with Majoranas (a Majorana island), however, these ground states partially split into two charge-parity sectors with the total number of electrons being even and odd respectively; this energy splitting is unrelated to Majorana mode hybridization, but comes from the charging energy and can be tuned by a gate voltage \cite{fu,xu}. This tunability enables electric control of Majoranas as well as new schemes of braiding  and quantum computation based on mesoscopic topological superconductor devices \cite{heck, hassler, grosfeld, alicea, hyart,terhal, vijay, landau}.

The interplay between Majorana modes and charging energy gives rise to a variety of topological quantum phenomena at the mesoscopic scale. One example is transport through a topological superconductor island with two spatially separated Majorana modes, each connected to a normal metal lead by electron tunneling~\cite{fu, egger, glazman}. Theory \cite{fu} predicts that an unusual resonant tunneling process involving two {\it distant} Majoranas gives rise to a phase-coherent charge-$e$ transport dubbed electron teleportation, exhibiting a conductance peak when the island is at a charge-degeneracy point.
In a recent groundbreaking experiment \cite{marcus} on proximitized nanowires under a magnetic field ---a promising platform for topological superconductivity \cite{sarma,oreg,kouwenhoven,weizmann}, $1e$-periodic zero-bias conductance through the superconducting island has been observed in the Coulomb blockade regime, providing experimental support for electron teleportation via Majorana modes.

In this work, we study multi-terminal charge transport through a Majorana island connected with $M>2$ leads, each tunnel coupled to a Majorana zero mode, as shown in Fig.~1. We assume these  Majoranas are far apart and have vanishing wavefunction hybridization. The charge on the island is  tuned by a gate voltage. This type of Majorana islands have recently been fabricated \cite{marcus2, marcus3} and attracted considerable interest.

Our study is also motivated by recent theoretical breakthroughs~\cite{ BeriCooper,AltlandEgger,Beri,Affleck13,AltlandBeryEggerTsvelik14,Tsvelik,numerics,ZazunovAltlandEgger14,Eriksson14a,Eriksson14,  Kashuba15,Pikulin16,Meidan16,Plugge16}, especially the seminal works of B{\'e}ri-Cooper~\cite{BeriCooper} and Altland-Egger~\cite{AltlandEgger}, predicting a ``topological Kondo effect'' in the Coulomb valley  regime where the charge of the topological superconductor island is fixed. Under this condition, the Majorana degrees of freedom are {\it constrained} to be in a given fermion parity sector and collectively form a $SO(N)$ impurity ``spin'', which interacts with bosonic excitations in the leads. Remarkably, this interaction gives rise to a non-Fermi-liquid fixed point without fine tuning. However, since the Kondo temperature is exponentially small, the intriguing phenomena associated with the topological Kondo fixed point is only accessible at very low temperature \cite{numerics}.

Our work focuses on charge transport in multi-terminal Majorana islands in the vicinity of the charge-degeneracy point, which until now has not been studied. At this point, the charge on the island  fluctuates between $N_0$ and $N_0+1$ as  electrons tunnel in and out of it. Consequently, the conductance at high temperature exhibits a Coulomb blockade peak on resonance, and the Majorana degrees of freedom are {\it unconstrained} but correlate  with the charge parity \cite{fu,xu}. Since charging energy permits only two charge states on the island,
 tunneling events at different leads are interrelated.

 As we show, due to high-order tunneling processes that build up quantum coherence, the system flows from the unstable weak-tunneling regime to the strong-coupling regime. We find  that
the strong-coupling limit of Majorana islands connected with  electron leads is described by a non-Fermi liquid fixed point, which is stable against gate voltage detuning  away from the charge-degeneracy point and anisotropy of tunnel couplings between the island and the leads. The zero-temperature conductance at this fixed point is {\it universal} and a {\it fraction} of the conductance quantum, %independent of the bare tunnel coupling and shows up as a plateau as a function of gate voltage. The multi-terminal conductance matrix
\beq
G_{ii} &=& -\frac{2(M-1)e^2}{Mh} \nonumber \\
G_{ij} &=&  \frac{2e^2}{M h}, \textrm{ for } i\neq j \label{conductance}
\eeq
where $G_{ij}$ relates the voltage on lead $j$ to the current in lead $i$ via the relation $I_{i} =\sum_{j=1}^M G_{ij} V_j$. Furthermore, the low-temperature correction to the conductance  has a power-law temperature dependence with a universal exponent $2(M-2)/M$. Importantly, at the charge-degeneracy point, the crossover from high-temperature Coulomb blockade regime to the universal conductance Eq.~\eqref{conductance} occurs at a temperature which is parametrically higher than the Kondo temperature in the Coulomb valley regime, see Fig.2. This greatly facilitates experimental observation of the non-Fermi liquid behavior and the universal conductance associated with electron teleportation in multi-terminal Majorana islands.

\begin{figure}[pt]
	\centering
	\includegraphics[scale=0.38]{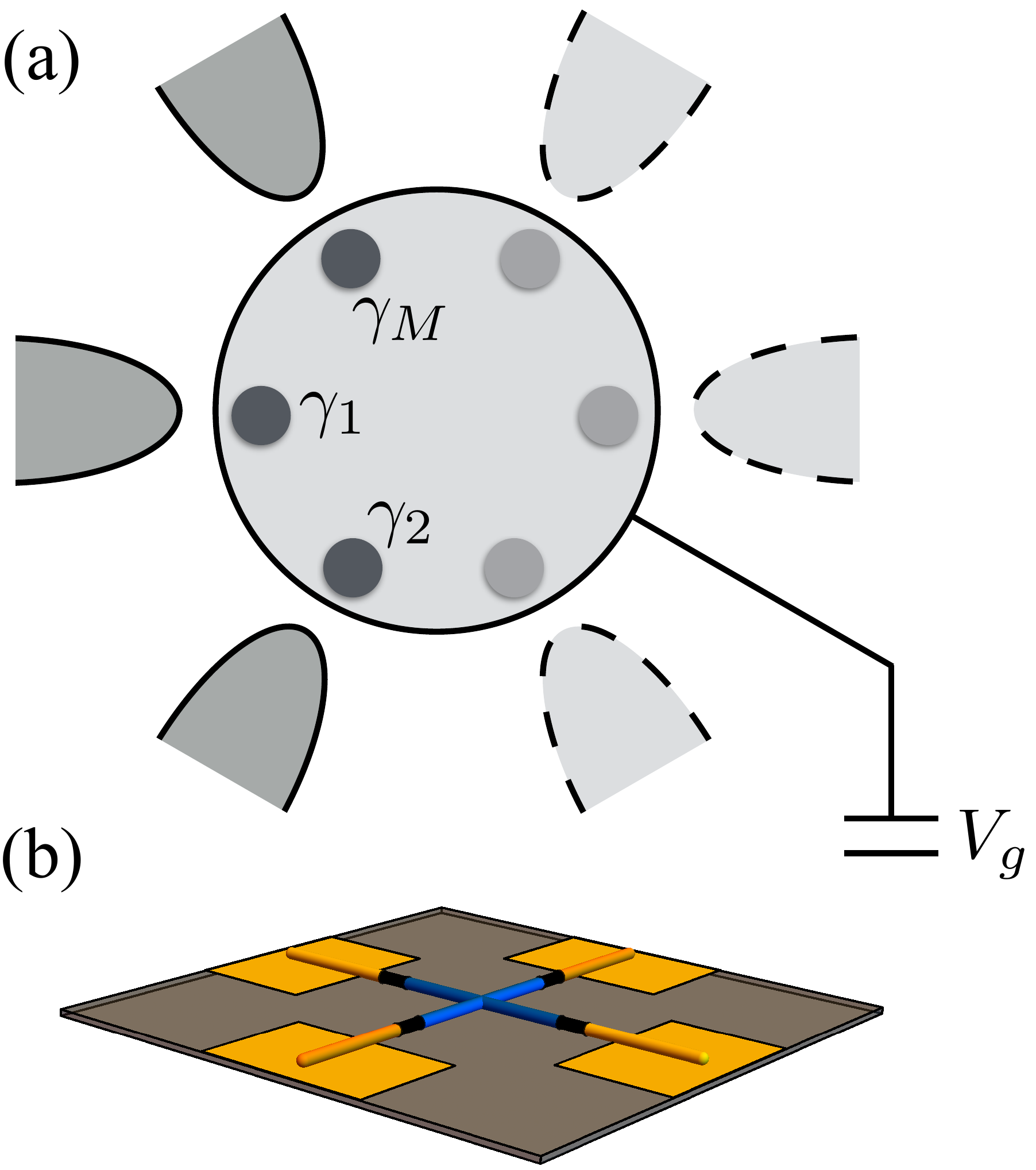}
	\label{fig1}
	\caption{(a) Device schematics: a topological superconductor island with spatially-separated Majorana modes  coupled to $M$ normal leads. A gate voltage $V_g$ tunes the charge on the island. (b) A  possible realization of our setup with $M=4$ using nanowires coated with a superconducting layer (blue).  An external magnetic field drives each proximitized wire into a topological superconductor phase hosting two Majorana mods at the ends, which are separated from the normal leads (orange) by a tunnel junction (black). }
\end{figure}

The Majorana nature of zero modes in the island is essential for the interesting physics described here. As we will show explicitly, Majoranas bind with  electrons in the leads to create a new type of emergent particle---a charge-$e$ boson, which governs conduction through the  island at low temperature.  Because of this transmutation from Fermi to Bose statistics, a Majorana island connected with {\it electron} leads becomes equivalent to a  particle interacting with {\it bosonic} reservoirs and undergoing quantum Brownian motion. This mapping then allows us to completely solve the problem of Majorana islands using a known strong-weak coupling duality \cite{YiKane}.

\textit{Model.--} Our multi-terminal Majorana island setup, shown in Fig.~1, is described by the Hamiltonian $H = H_{\rm{leads}} + H_{\rm{island}}+H_{\rm{T}}$. %The leads are assumed to be spinless with density of states $\rho$ at the Fermi level.
The superconducting island is capacitively coupled to a gate which determines its charging energy $E_c$ and average occupancy $n_g$ as
\begin{align}
H_{\rm{island}} = E_c \left(\hat{N}-n_g\right)^2.
\end{align}
Here $\hat N$ is the electron number operator of the island.  %conjugate to the superconducting phase $\theta$.
Importantly, due to the presence of zero-energy Majorana modes, the topological superconductor island admits an odd number of electrons on equal footing with an even number of electrons, without paying the energy cost of the superconducting gap (which is assumed to be the largest energy scale). Hence, the electron number $N$ is allowed to be either even or odd.
%, %and the Cooper pair operator $e^{\pm i\theta}$ increases/dereases $N$ by 2.
%and has the commutation relation with the phase:  $[\theta,\hat{N}]=2i$ \cite{fu}.

The island is coupled to the leads via single-electron tunneling described by~\cite{fu}
\begin{align}
H_{\rm{T}} =\sum_{j=1}^M t_j \psi_j^\dagger(0) \gamma_j e^{-i \hat \theta/2}+{\rm{H.c.}}, \label{HT}
\end{align}
where $\psi^\dagger_j(0)$ creates an electron at the end of lead $j$.
$\gamma_1, ..., \gamma_M$ are Majorana mode operators with the defining property
\begin{align}
\gamma_j^\dagger = \gamma_j, \; \{ \gamma_i ,\gamma_j \} = 2 \delta_{ij}. \label{gamma}
\end{align}
%$\gamma_j=\int{dx}\xi_j(x)e^{-i\theta/2}f_j^{\dag}+\rm{H.c.}$, satisfying Clifford anticommutator algebra, $\{ \gamma_i ,\gamma_j \} = 2 \delta_{ij}$, are Majorana operators expressed in terms of the wave functions $\xi_j(x)$ of bound states inside the island localized near the $j$-th lead.
These Majorana modes are assumed to be far apart without direct coupling. The superconducting phase $\hat \theta$ is conjugate to the electron number $\hat N$, with the commutation relation
$[\hat \theta,\hat{N}]=2i$, so that $e^{\pm i \hat \theta/2}$ changes the number of electrons in the island by $\pm 1$. As a single electron tunnels in (out of) the island from (to) the leads, the tunneling operator Eq.~\eqref{HT} simultaneously flips the fermion parity of the island---which is encoded in Majorana degrees of freedom---and changes the charge on the island by $\pm e$.

We specialize to the case where $E_c$ dominates over both the temperature $T$ and the level broadening induced by coupling to leads $\Gamma = \sum_j \Gamma_j =\sum_j \rho {t}_j^2$, where $\rho$ is the density of states at the leads.
Then, for the range of gate voltages corresponding to $N_0 < n_g < N_0 +1$, only two charge states with $N=N_0$ and $N_0+1$ are relevant at low energy. We denote these two charge states by a pseudo-spin $\sigma^z=\mp 1$, and project the full Hamiltonian $H$ to the low-energy Hilbert space to obtain
\begin{equation}
H =H_{\rm{leads}}+ \sum_{j=1}^M \left( t_j \psi_j^\dagger \sigma^- \gamma_j+{\rm{H.c.}} \right) + \Delta_g \sigma^z, \label{Heff}
\end{equation}
where $2\Delta_g \equiv 2E_c(N_0-n_g + 1/2)$ is the energy difference of the two charge states.

At high temperatures (yet lower than ${E}_c$), the conductance through the Majorana island exhibits a resonance peak as the gate voltage is swept across  the charge-degeneracy point $\Delta_g=0$. Near this point and to  leading order in tunnel coupling, the conductance peak is described by conventional sequential tunneling through an impurity level~\cite{heck}:
\begin{equation}
\label{eq:Coulombpaks}
G_{ij}  = \frac{e^2}{h}\frac{ \Gamma_i \left( \Gamma_j/\Gamma - \delta_{ij} \right)}{4T\cosh^2(\Delta_g/T)}.
\end{equation}
Coherent tunneling processes  due to the Majorana modes manifest themselves in higher-order corrections in $\Gamma/T$, and thus the crossover into the strong-coupling limit occurs at $T \sim T^{*} \equiv \Gamma$ (see Fig.~2).

\textit{Statistical Transmutation.--} To obtain the multi-terminal conductance at low temperature requires a non-perturbative strong-coupling analysis. First, without loss of generality, we model the noninteracting electrons in the leads as chiral fermions moving in infinite one-dimensional wires:
\begin{align}\label{HleadsChiral}
H_{\rm{leads}} = \frac{1}{2\pi}\sum_{j=1}^M \int_{-\infty}^\infty dx \; v   \psi_{j}^\dagger i \partial_x \psi_{j},  %- \psi_{j,O}^\dagger \partial_x \psi_{j,O}
\end{align}
where $\psi_j^\dagger$  at different leads anticommute,
$\{ \psi^\dagger_i(x), \psi^\dagger_j(x') \} =\{\psi^\dagger_i(x), \psi_j (x') \} = 0$ for $i \neq j$.

We note that the tunneling operator shown in Eqs.~\eqref{HT} and \eqref{Heff} involves a product of an electron operator ($\psi^\dagger_j$ or $\psi_j$) and the self-adjoint Majorana operator $\gamma_j$. Such bilinear operators defined at different leads are {\it bosonic} and {\it mutually-commuting},
\begin{align}
[ \psi^\dagger_i (x) \gamma_i, \psi^\dagger_j (x') \gamma_j ] =
[ \psi^\dagger_i (x) \gamma_i, \psi_j (x') \gamma_j ]=0,
%= [ \psi^\dagger_i \gamma_i, \psi_j \gamma_j ]=0
\end{align}
for $i \neq j$.
This all-commuting condition allows us to bosonize $\{ \psi^\dagger_j (x) \gamma_j \}$ into $M$ independent chiral boson fields:
\begin{align}
\psi^\dagger_j(x) \gamma_j \sim e^{i \varphi_j(x)} , \; j=1,...,M.
\end{align}
Details of this bosonization procedure can be found in the appendix.

After bosonization the imaginary-time action describing the leads is given by:
\begin{equation}
%S_{\rm{lead}} = \frac{1}{2 \pi} \sum_{j=1}^M \int_{-\infty}^0 dx \int d \tau [(\partial_\tau \varphi_j)^2+v^2(\partial_x \varphi_j)^2]	,	
S_{\rm{leads}} =\frac{1}{4 \pi}\sum_{j=1}^{M} \int_{-\infty}^\infty dx \int d \tau  \partial_x \varphi_j(v\partial_x \varphi_j - i \partial_\tau \varphi_j),
\label{bosonized-lead}
\end{equation}
and the tunneling term at $x=0$ becomes
\begin{equation}
\label{Stun}
S_{\rm{T}}=  \sum_{j=1}^M \int d \tau  \; t_j  e^{i \varphi_j(0,\tau)}	\sigma^- + {\rm{H.c.}}.
\end{equation}
%The charging energy term  becomes
%\begin{equation}
%$S_{\rm{island}} = \Delta_g \sigma^z$.

%\end{equation}

\begin{figure}[ht]
	\centering	
	\includegraphics[scale=0.3]{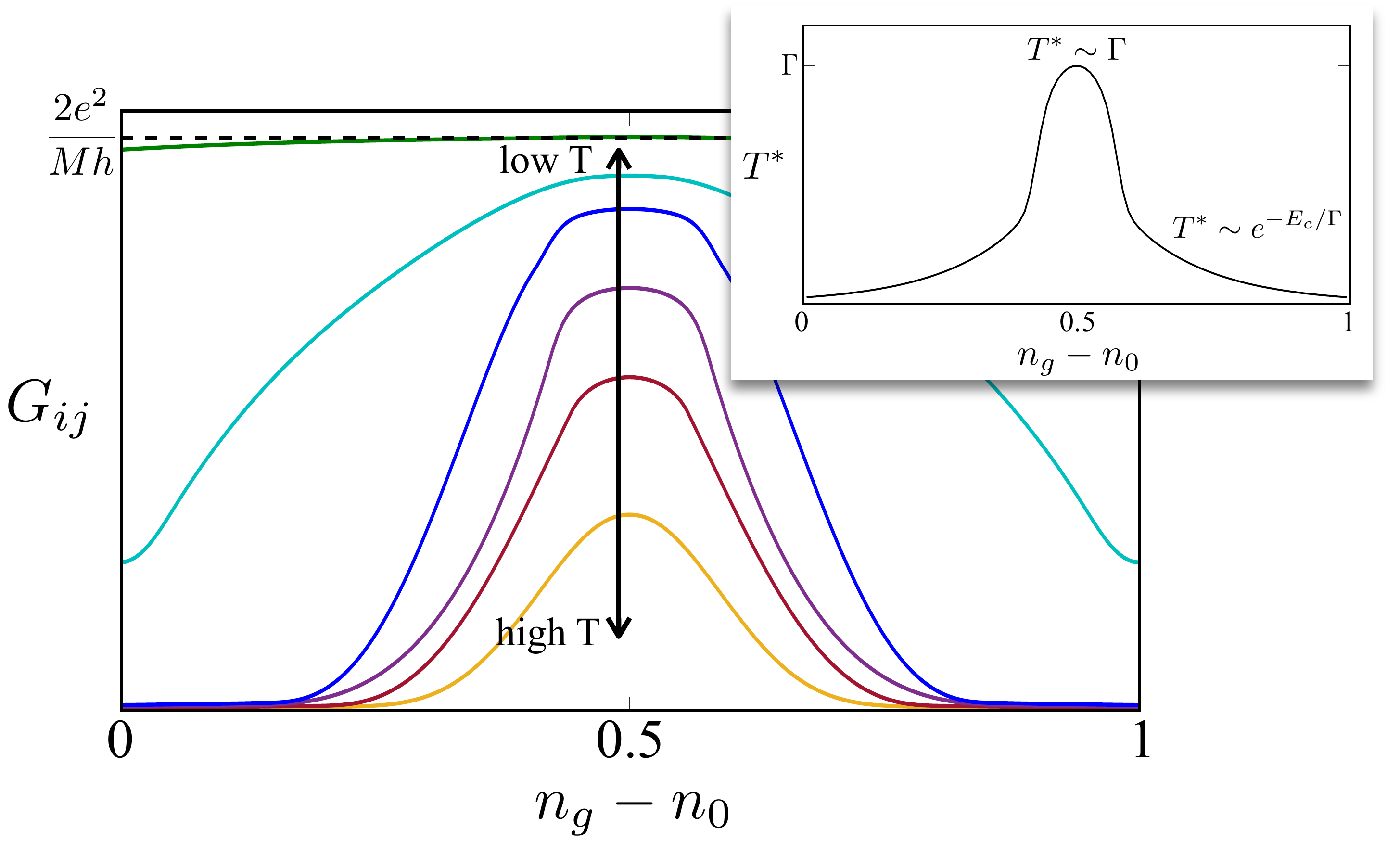}
\label{wignerfig}
	\caption{The conductance of multi-terminal Majorana islands ($M\geq3$) between any two normal leads as a function of gate voltage is plotted for various temperatures. %between $T=\Gamma$ (yellow curve) and $T\rightarrow0$ (black curve).
At high temperature, the conductance shows a Coulomb blockade peak near the charge-degeneracy point $n_g=1/2$, see Eq.~\eqref{eq:Coulombpaks}.  At low temperatures $T\ll T^{*}$ the conductance takes a universal form, Eq.~\eqref{GlowT}, and approaches the universal conductance $\frac{2e^2}{Mh}$ at $T=0$. Conductance curves at intermediate temperatures are interpolations between the two limits. The inset shows the strong dependence of the crossover temperature $T^{*}$ on the gate voltage: it is maximal and of the order of the level broadening $\Gamma$ at the Coulomb peak
($n_g=1/2$), and becomes exponentially small in the Coulomb valley, corresponding to the Kondo temperature.}
\end{figure}

We have thus made an exact transformation mapping the problem of {\it electron} tunneling between a Majorana island and leads to a problem of {\it boson} hopping between an impurity level and reservoirs. This transformation is enabled by the presence of Majorana modes, which bind with electrons in the leads to form charge-$e$ bosons. To appreciate the importance of Majorana-enabled statistical transmutation, it is instructive to compare and contrast teleportation through Majorana islands with resonant tunneling through a single-particle energy level in a quantum dot. In both cases, the charge on the island or the dot fluctuates between two values differing by $1e$. Consider the sequence of successive tunneling events shown in Fig.~3 that exchanges two electrons on lead 1 and 3 via lead 2. The amplitude of this exchange process
in a perturbative expansion in powers of the tunneling operator is negative for resonant tunneling in a quantum dot as expected for such free fermion problem. However, by a straightforward calculation  using Eq.~\eqref{Heff}, one finds this amplitude is positive for teleportation in Majorana islands, showing that the effective charge carrier here is a boson. This comparison explains why the bosonized action for Majorana islands, Eq.~\eqref{Stun}, does not apply to resonant tunneling through an energy level; the latter problem involves Klein factors necessary for keeping track of electron's Fermi statistics \cite{Nayak}. We note that exchange processes are present only for setups with more than two leads. Therefore, electron teleportation in two-terminal Majorana islands \cite{fu} is a special case where the effect of statistical transmutation is nulled.

 \begin{figure}[pt]
	\centering
	\includegraphics[scale=0.3]{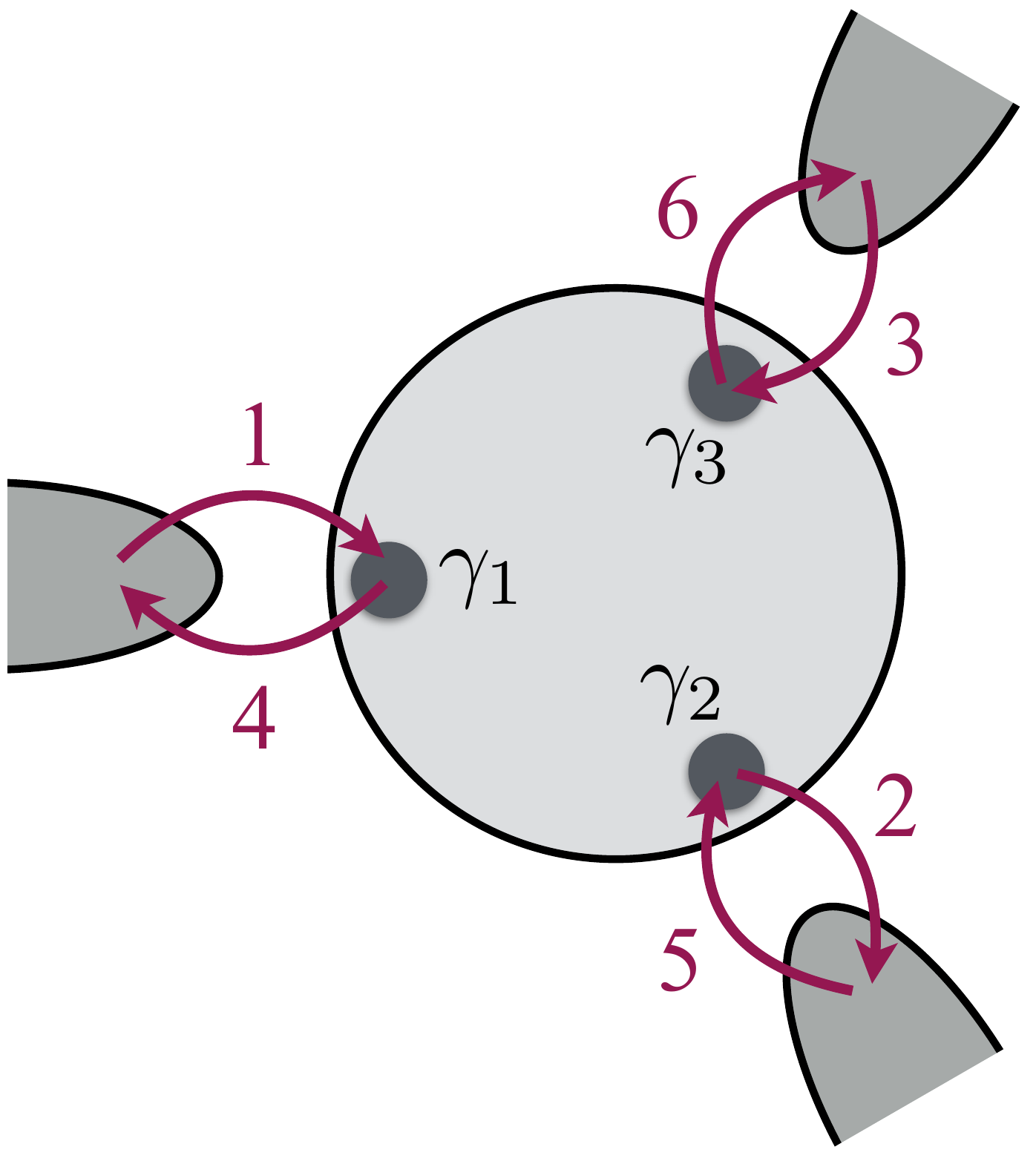}
	\caption{A sequence of six successive tunneling events that exchanges two electrons on leads 1 and 3 via lead 2. The amplitude of this process is positive for electron teleportation in a  Majorana island, unlike the negative sign for resonant tunneling into a single-particle state. This sign change demonstrates that the effective charge carrier is bosonic rather than fermionic. }
\end{figure}

%Similar looking problem of resonant multi-lead tunneling via a level was studied in Ref.~\onlinecite{Nayak} and its relation to the quantum brownian motion (QBM) on a periodic potential Yi and Kane~\cite{YiKane} has been the subject of many discussions; see~\cite{Chamon} and references therein. However, it was noticed that the (fermionic) Klein factors $\eta_i$ in usual tunneling junctions lead to additional minus signs as compared to the QBM model which is bosonic in nature and change the low energy physics.

\textit{Mapping to quantum Brownian motion.--}
We start the strong-coupling analysis by studying  Majorana islands at the charge-degeneracy point $\Delta_g=0$ and with equal tunnel couplings to all leads: $t_1= t_2 = ...=t_M \equiv J_\perp$. The bosonized action in Eqs.~\eqref{bosonized-lead} and \eqref{Stun} is then equivalent to the action of quantum Brownian motion (QBM) of a particle in a  periodic potential, as shown by Yi and Kane \cite{YiKane}. To see this mapping, we  integrate out the degrees of freedom away from $x=0$ in the leads to  obtain a $0+1$-dimensional action in terms of the boson phase fields $(\varphi_1, ..., \varphi_M)|_{x=0}\equiv {\vec \varphi}$, given by $S=S_0 + S_{T}$ where
 \beq\label{Dissipation}
 S_0 =\frac{1}{(2 \pi)^2} \int d \omega |\omega| |\vec \varphi(\omega)|^2,
 \eeq
 describes the leads,
 and
 \beq
 S_T= J_\perp \sum_j e^{i \sqrt{2} {\vec \varphi} \cdot {\vec R}_0^{(j)}} \sigma^- + {\rm H.c}, \label{hop}
 \eeq
describes the tunneling between the leads and the island. Here
${\vec R}_0^{(j)}$ is a $M$-dimensional vector whose $j$-th component is $\frac{1}{\sqrt{2}}$ and other components are all zero, so that $ \vec \varphi \cdot {\vec R}_0^{(j)} = \varphi_j(0)/\sqrt{2}$. We have included the normalization factor $\sqrt{2}$ in  Eq.~\eqref{hop} so that the scaling dimension of $S_T$ is equal to $|R_0|^2=\frac{1}{2}$  (for more details see Appendix).

We now identify $\vec \varphi$ as the momentum of a particle coupled to a dissipative bath. The number of charges carriers in the leads $(n_1,...,n_M)$---which is conjugate to $\vec \varphi$---corresponds to  the particle's coordinate $\vec r$. For small $J_{\perp}$, the action $S$ describes QBM of this particle in a strong periodic potential, whose minima are located at $\sum_{j=1}^M n_j \cdot (\sqrt{2} \vec{R}_0^{(j)})$. Specifically, $S_0$ determines the amount of dissipation, and $S_T$, being a translation operator, generates a small probability of particle hopping between two adjacent potential minima connected by the lattice vector $\vec R_0$.  %(e.g. $n_1,n_2,n_3$ for $M=3$, see Fig.~3)

In our setup, the sum of all charges on the leads $\sum_j n_j$ may only fluctuate by 1 due to charge conservation and the restriction of two allowed charge states $N_0$ and $N_0+1$ on the island. This implies that the Brownian particle is only allowed to hop on two adjacent lattice planes perpendicular to the direction $\hat{R}_\perp=\frac{1}{\sqrt{M}}(1,1,...,1)$. For $M=3$, the potential minima of the Brownian particle form a corrugated honeycomb lattice, consisting of two triangular sublattices, as illustarted in Fig.~4. For $M>3$, the particle hops on the generalization of corrugated honeycomb lattices in $M-1$ dimensions. The two sublattices correspond to $\sigma^z=\pm 1$, hence the particle hopping described by Eq.~\eqref{hop} alternates between the two sublattices.

For noninteracting electron leads, the hopping operator Eq.~\eqref{hop} has a scaling dimension given by $|R_0|^2=1/2<1$, which is relevant at the disconnected fixed point $J_\perp=0$. Thus the strong potential limit of QBM, described in terms of particle hopping between deep potential minima, is unstable and flows under  RG to a different fixed point.

\begin{figure}[pt]
	\centering
	\includegraphics[scale=0.3]{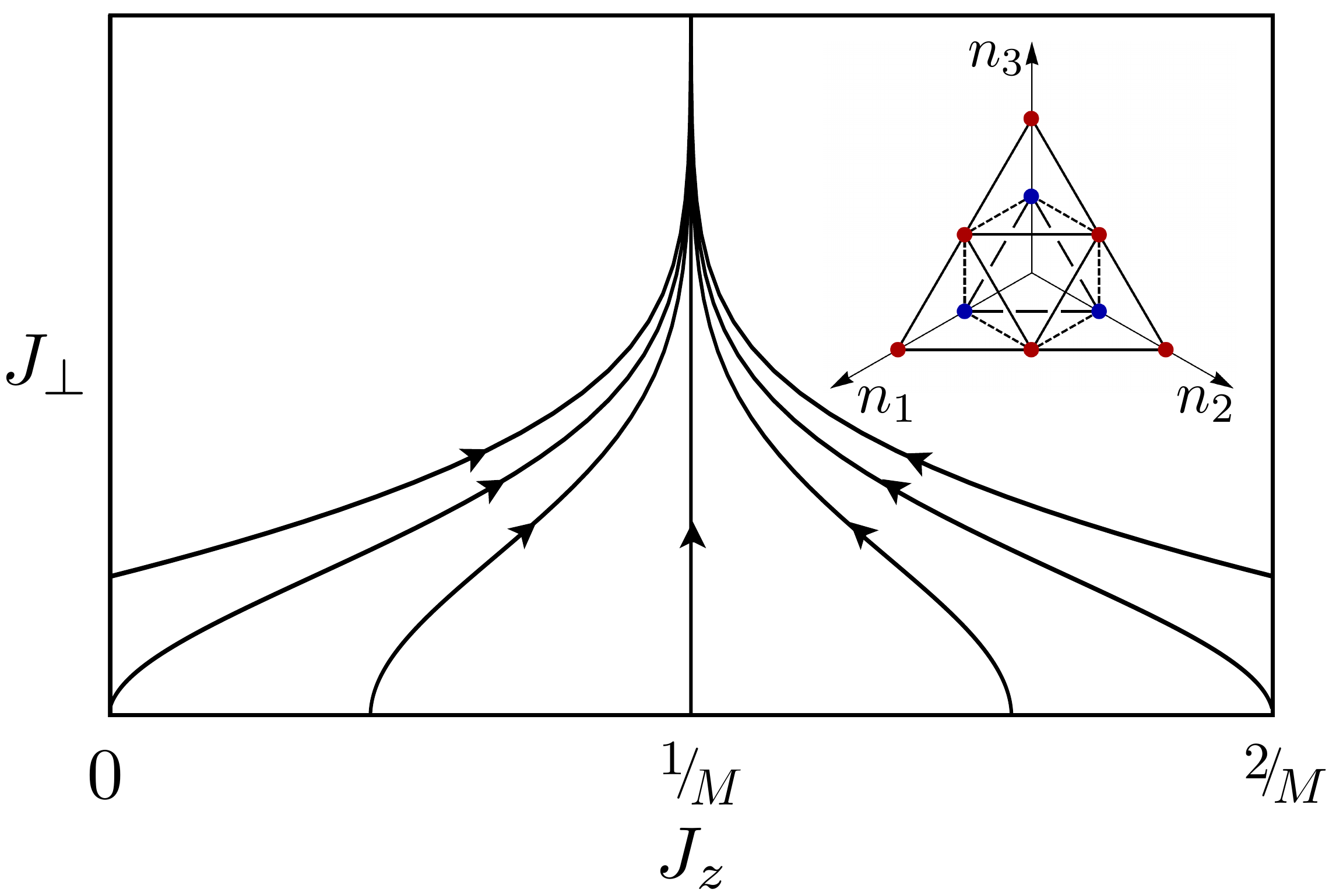}
	\caption{Renormalization group (RG) flow  in the weak-coupling regime showing a Toulouse-like limit at $J_z^*=1/M$. There, the bosonized action for electron teleportation in Majorana islands Eq.~\eqref{st} is equivalent to quantum Brownian motion on the $M-1$ dimensional honeycomb lattice (shown for $M=3$ in the inset), whose sites correspond to  allowed charge configurations of the leads. %There, the distance between the two  sublattices corresponding to $\sigma_z=\mp1$ vanishes, and the QBM is confined to a $M-1$-dimensional lattice.
}
\end{figure}

To identify this new fixed point,
we first note that before integrating out degrees of freedom in the leads, a new term $\frac{v}{2}J_z \sigma^z \sum_j \partial_x \varphi_j(x) \delta(x)$ is generated in the RG process.
The perturbative RG equations for the two coupling constants $J_\perp$ and $J_z$ are:
\begin{eqnarray}
\label{RGTouloyes}
\frac{dJ_z}{ d l} &=& J_\perp^2(1-M J_z),\nonumber \\
\frac{dJ_\perp}{ d l} &=& \frac{1}{2} J_\perp+J_\perp J_z [1-(M/2) J_z].
\end{eqnarray}
The resulting RG flow, plotted in Fig.~4, shows that $J_z$, even with initial value of $0$, flows to a Toulouse-like  point in which $J_z^*=1/M$. At this point, a unitary transformation $U = e^{i \sigma^z \sum_j \varphi_j (0)/(2M)}$ eliminates the $J_z$ term from the action, similar to the analysis of Ref.~\cite{YiKane}. After performing the transformation, the hopping operator,  Eq.~\eqref{hop}, becomes
\begin{eqnarray}
S^*_T = J_\perp \sum_{j=1}^M e^{ i \sqrt{2} \vec{\varphi} \cdot \vec{R}^{(j)}_\parallel} \sigma^- + {\rm{H.c.}}, \label{st}
\end{eqnarray}
where the vectors $\vec{R}^{(1)}_\parallel, ..., \vec{R}^{(M)}_\parallel $ are all orthogonal to $\hat R_\perp=\frac{1}{\sqrt{M}}(1,1,1..)$ and have the length $|R_\parallel| = \sqrt{1-1/M} |R_0|$. Importantly, the total charge field $\varphi_c = \vec{\varphi} \cdot {\hat R}_\perp$ which corresponds to the motion of the Brownian particle along the $\hat R_\perp$ direction, disappears from $S_T^*$. As a result, the motion along $\hat{R}_{\perp}$ is decoupled from the motion in the perpendicular direction, which is spanned by the remaining  $M-1$ linearly independent vectors appearing in $S_T^*$.  Therefore, independent of the bare coupling constant $J_z$, the system flows to the Toulouse limit with an action $S_0 + S_T^*$ that is equivalent to QBM on a $(M-1)$-dimensional honeycomb lattice.

%Yi and Kane have constructed the phase diagram in such a lattice~\cite{YiKane}. The scaling dimension of the tunneling operator $J_\perp$, dictating the flow either to the weak or strong barrier limit, differs from its initial value at $J_z=0$: initially, at $J_z=0$, we had $R_\perp^2=\frac{1}{2M}$. The scaling dimension of the original tunneling electron becomes $|R_\parallel|^2 + R_\perp^2 \to |R_\parallel|^2 = \frac{1}{2} - \frac{1}{2M}$. For $M=3$, for example, it is $\frac{1}{3}$.

%We now argue that the fixed point of the weak triangular lattice is identical to the one obtained in the honeycomb lattice.  To show this we first treat the corrugation of the initial honeycomb lattice.

The hopping between deep potential minima of the honeycomb lattice has a scaling dimension given by $|R_\parallel|^2$, which is smaller than $1$ for all $M$, and thus is a relevant operator. As a result, the hopping amplitude $J_{\perp}$ grows, or equivalently the periodic potential weakens in the RG process. Next, we consider the limit of vanishing periodic potential, or QBM in free space.  We analyze its stability against applying a periodic potential with the same periodicity as the original honeycomb lattice ~\cite{YiKane}. Such a potential can be decomposed into Fourier components: $U(\vec r)=\sum_{\vec G} v_{\vec G} e^{i \vec G \cdot \vec r}$, where $\vec G$ is the reciprocal lattice vector defined by $\vec G \cdot \vec R = {\rm{integer}}$ for any Bravais lattice vector $\vec{R}$ of the honeycomb lattice. %The shortest $\vec{R}$ includes e.g., $(R_{\parallel}^{(1)}-R_{\parallel}^{(2)})_i=(\delta_{i1}-\delta_{i2})/\sqrt{2}$.}
The scaling dimension of the $v_{\vec G}$ component of the perturbation is given by $|\vec G |^2$ (see Appendix). The shortest reciprocal-lattice vector $\vec G_0$ is of length $\sqrt{2(1-1/M)}$.  Therefore,  the periodic potential $U(\vec r)$ is marginal for $M=2$, and irrelevant for $M>2$. As argued by Yi and Kane ~\cite{YiKane}, the contrasting stability in the limit of strong and weak potential $U(\vec{r})$ implies that the periodic potential flows to zero in the RG process, leading to QBM in free space as the infrared fixed point.

We now turn to Majorana islands detuned away from the charge-degeneracy point and/or having unequal coupling to the leads. In the QBM formulation, the deviation from $\Delta_g \sigma^z=0$ makes the two sublattices of the honeycomb lattice inequivalent. Unequal tunnel couplings described by $\sum_j \delta_j e^{i \varphi_j(0)} \sigma^- + {\rm H.c.}$ make the honeycomb lattice spatially anisotropic. Both perturbations correspond to deformations of the honeycomb lattice that lower its crystal symmetry but does not alter its periodicity. As such, they are irrelevant at the free QBM fixed point as shown by our stability analysis. We thus conclude that the strong-coupling limit of Majorana islands connected with $M>2$ noninteracting electron leads is a non-Fermi liquid fixed point that maps to QBM in free space and is stable against asymmetric coupling to leads and gate voltage detuning away from the charge-degeneracy point.

\textit{Universal conductance and low-temperature corrections.}
The isotropy of QBM in free space implies that at the infrared fixed point, all off-diagonal components of the multi-terminal conductance matrix are equal: $G_{ij}=G_0$ for $i\neq j$. Current conservation then implies that $G_{ii} = -\sum_{j\neq i} G_{ij} = -(M-1) G_0 $. To determine $G_0$, let us consider the following setup: we apply a voltage $V_1 = V/2$ on the first lead, $V_2 = -V/2$ on the second lead, and $V_j=0$ for all other leads. By definition, the resulting current is
\begin{eqnarray}\label{Current}
I_2= \sum_{j=1}^{M}G_{2j}V_j = M G_0 V/2=-I_1,
\end{eqnarray}
while $I_j=0$ in all other leads. Finding the current $I_{1,2}$ for this particular voltage setup will then yield $G_0$, thus the entire matrix $G_{ij}$. %\sout{To calculate the current, note that $V_j$ couples linearly to the number of charges in the lead $n_j$, which corresponds to the coordinate of the Brownian particle {\cb{$r_j$}}. Thus $V_1=-V_2 = V/2$ adds a perturbation to the action $S_0$ of QBM in free space, {\cb{$\propto \frac{a(t)}{2} \partial_t (r_1 - r_2)$}}, {\cb{[[Remark: there is a $2\pi/\sqrt{2}$ proportionality between $n_j$ and $r_j=\theta_j$ in our conventions, SM, but anyway the prefactor is not important in this argument]]}}  where in real time $V_j(t)=\partial_t a_j(t)$.} \KM{[a bit destroying the flow and not needed here, I think it should appear later in the text]}
The voltage $V_j$ couples to the charges on lead $n_j$, and hence corresponds to adding a linear potential to the coordinate of the Brownian particle $r_j$. The uniform force field in the direction $(1,-1,0...0)$ and the coupling to the dissipative bath, give rise to a nonzero steady-state velocity in this direction. Since QBM in free space is spatially isotropic and direction independent, the steady-state velocity is independent of spatial dimensionality $M-1$, and hence so is the current $I_{1,2}$. For $M=2$ it was shown~\cite{fu} that a Majorana island with equal tunnel couplings to two leads maps to resonant electron tunneling, for which  $I_2 = -I_1 = \frac{e^2}{h}V$. Therefore, equating the known result for $M=2$ and Eq.~\eqref{Current} we obtain $G_0=\frac{2 e^2}{M h}$ for all $M$, yielding the universal multi-terminal conductance in Eq.~\eqref{conductance}. It is interesting to note that in the limit $M\rightarrow \infty$, the conductance $G_{ii}$ which determines the total current through the island approaches $2e^2/h$, which is identical to the conductance from resonant Andreev reflection from a single  Majorana mode in a {\it grounded} superconductor. This is consistent with the expectation that coupling the island to a large number of leads  makes it effectively grounded.

%\KM{We emphasize that the derivation of the low-temperature conductance is equivalent to writing the action for {$r_j$}, and coupling it to a vector potential $i\frac{a(\tau)}{2} \partial_\tau (r_1 - r_2)$ where $V_j(\tau)=i\partial_\tau a_j(\tau)$. Then, the conductance can be found by calculating the response to $a$.  At the fixed point, the response is fully diamagnetic, while at finite temperature corrections due to the weak periodic potential emerge~\cite{KaneFisher}.  }

At finite but low temperature, corrections to the conductance are governed by the leading irrelevant operator at the infrared fixed point. In the QBM formulation, this operator corresponds to adding a weak honeycomb potential, which has the scaling dimension $\Delta_M =  \frac{2(M-1)}{M}$.
%c.f. Ref.~\onlinecite{Eriksson}.
This gives rise to a universal power-law correction to the conductance at low temperature,
\begin{equation}
\label{GlowT}
G_{i\neq j} =\frac{e^2}{ h} \left[  \frac{2}{M } - c\left( \frac{T}{T^*} \right)^{2 \Delta_M - 2} \right],
\end{equation}
where $c$ is a constant of order $1$.  The temperature $T^*$ depends strongly on the gate voltage:  %Note that the although the low-temperature conductance has a universal form, it depends on the gate voltage through $T^{*}$. As demonstrated in Fig.~2,
near the charge-degeneracy point $T^{*}\sim\Gamma$ is significantly higher than in the Kondo regime $\sim{e}^{-E_c/\Gamma}$. Consequently, coherence effects become important at higher temperatures for $\Delta_g\approx0$, and the conductance approaches its zero-temperature universal value faster (see Fig.~2).

\textit{Kondo regime--} When the gate voltage is tuned to the Coulomb valley ($\Delta_g \gg T, \Gamma$), the charge on the island is fixed to an integer $N_0$. As a result, electrons can no longer hop into or out of the island. Instead, virtual tunneling processes give rise to an effective exchange interaction that transfers charge between the leads while switching the state of Majoranas within a fermion parity sector given by $N_0 \mod 2$. This Kondo-type interaction $H_{\rm K}$ can also be derived from our model, Eq.~\eqref{Heff}, via second-order perturbation theory in $t_i$, which yields \cite{BeriCooper}
\begin{eqnarray}
H_{\rm K} =  \sum_{i \ne j}^M \lambda_{ij} (\psi_i^\dagger \psi_j  - \psi_j^\dagger \psi_i) O_{ij}, \label{HK}
\end{eqnarray}
where $O_{ij}=\gamma_i \gamma_j$ are $SO(M)$ generators satisfying the Clifford algebra, and the Kondo coupling is $\lambda_{ij} \propto t_i t_j /\Delta_g $.

%Clearly, $H_K$ has a completely different form as the

\begin{figure}[pt]
	\centering
	\includegraphics[scale=0.4]{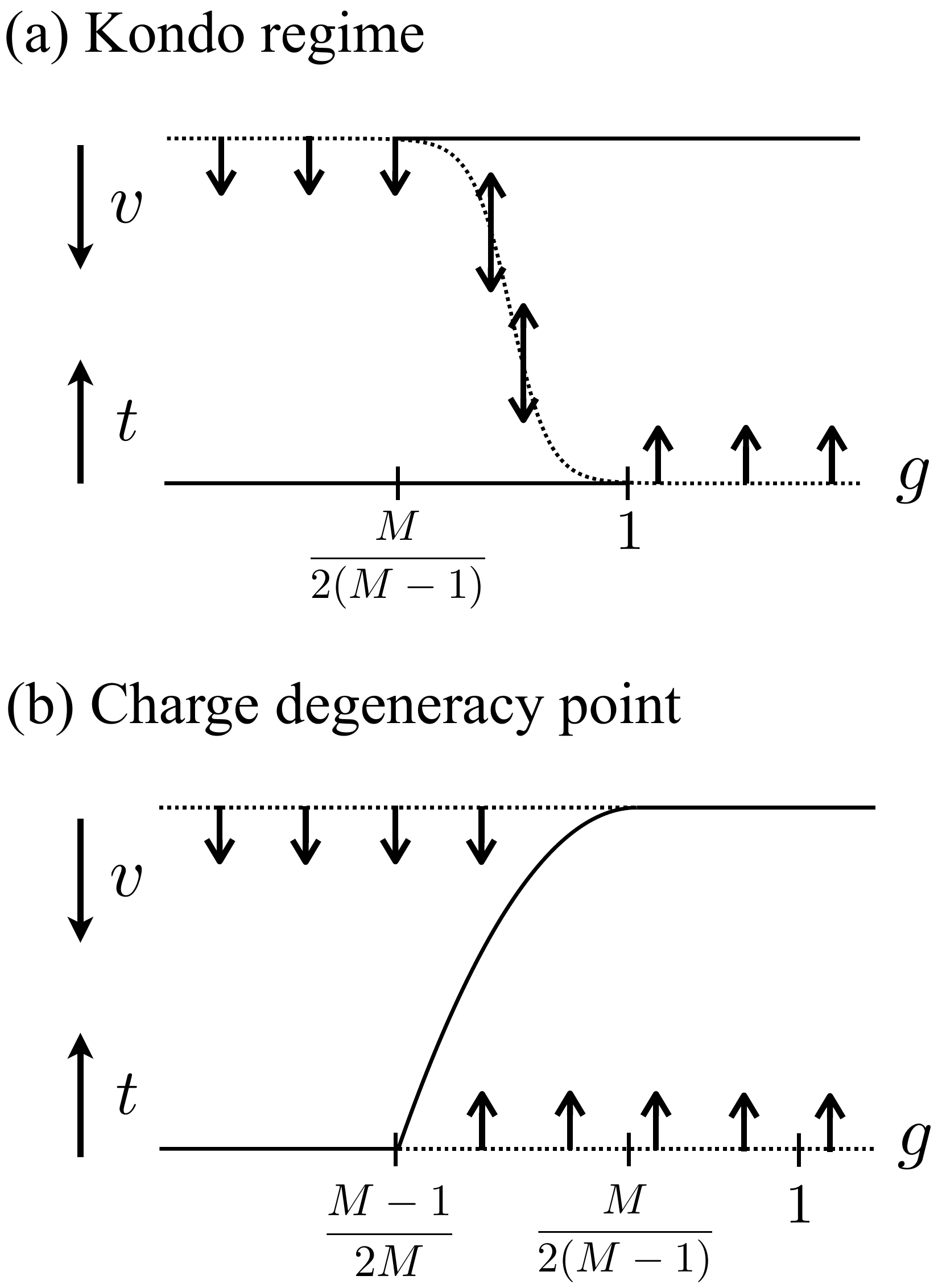}
	\label{wignerfig1}
	\caption{Phase diagrams and RG flows for a Majorana island connected to $M$ electron leads, in the Kondo regime (a) and near the charge-degeneracy point (b), as a function of
the Luttinger parameter $g$ characterizing the strength of interaction in the leads ($g=1$ for noninteracting leads).
In both (a) and (b), stable (unstable) fixed points are marked by solid (dashed) lines. The lower line corresponds to the weak coupling limit of the island and the leads ($t\rightarrow0$). The upper line corresponds to the strong-coupling limit that maps to quantum Brownian motion in a weak periodic potential ($v\rightarrow0$). %where the leading operator has a scaling dimension $\Delta_M$ given by Eq.xx and xx.
In the Kondo regime, the weak coupling limit is stable for arbitrarily weak repulsive interactions $g<1$. In contrast, near the charge-degeneracy point this limit is unstable: it flows to the strong-coupling fixed point for $\frac{M}{2(M-1)}<g<1$, and to a stable intermediate fixed point for $\frac{M-1}{2M}<g<\frac{M}{2(M-1)}$. These flow diagrams are generalizations of the result of Ref.~\cite{YiKane} for $M=3$ to all $M$.}
\end{figure}

As shown by  B{\'e}ri \cite{Beri}, this Kondo problem of bosonic nature directly maps to QBM on a triangular lattice. This mapping can also be understood in our formulation: a large $\Delta_g$ adds a strong sublattice potential to the corrugated honeycomb lattice, so that the Brownian particle hops between sites on the low-energy sublattice via virtual transitions through the high-energy sublattice. Importantly, the hopping operator $H_K$ in the Kondo regime is marginally relevant and sets the length of the triangular lattice vector to be $|\vec{R}|=1$.
%, different from  $|R_0|=1/\sqrt{2}$ set by the relevant hopping operator $S_T$ at the charge-degeneracy point.
Analysis of Majorana islands in the Kondo regime~\cite{Beri} reveals that for noninteracting leads the strong-coupling fixed point also maps to QBM in free space,  which is the same as the fixed point we found in the vicinity of $\Delta_g=0$ (see also Appendix). Thus, we conclude that despite having significantly different conductance in the high-temperature Coulomb peak and Coulomb valley regime, the system exhibits the universal conductance Eq.~\eqref{conductance} at $T=0$,
independent of the gate voltage. Our result generalizes Ref.\cite{Beri} where the $T=0$ conductance was found in the Kondo regime.
However, as we show below, the Coulomb peak and Coulomb valley regime of a Majorana island flow to different infrared fixed points for repulsive interactions in the leads.

\textit{Interacting leads--}
The QBM formulation provides a unified framework for analyzing Majorana islands both in the vicinity of the charge-degeneracy point and in the Kondo regime. Although up to here we have considered noninteracting electron leads, the generalization to the interacting case is straightforward.  To study interaction effects, we only need to identify the change in the lengths of the direct and reciprocal lattice vectors, given by $|R_0|\rightarrow |R_0|/\sqrt{g}$ and $|G|\rightarrow\sqrt{g}{|G|}$ , where $g$ is the Luttinger parameter~\cite{YiKane}. Therefore, in the Kondo regime ($|\vec{R}|=1/\sqrt{g}$),  arbitrarily weak repulsive interactions $g<1$ make $H_K$ irrelevant, so that the limit of decoupled Majorana island and leads is stable against weak tunnel couplings. As the couplings $\lambda_{i,j}$ increase above a critical value, the system undergoes a quantum phase transition~\cite{AltlandEgger,Beri} into the strong-coupling fixed point (see Fig.~5).

In contrast, near the charge-degeneracy point,  electron tunneling into the Majorana island $S_{\rm T}$ remains a relevant operator over a finite range of interaction strengths. Since  $|\vec{R}_{||}|=\sqrt{(M-1)/(2Mg)}$ and $|\vec{G}|= \sqrt{2g(M-1)/M}$, we obtain that for  $g>\frac{M}{2(M-1)}$, %no new relevant operators appear and
the system flows from the unstable weak-tunneling to the stable strong-coupling fixed point.  %\KM{(characterized by a $g$-dependent conductance, $G_{ij}=g\frac{2e^2}{Mh}$)}.  %However, the low temperature conductance  is modified by interactions  $G_{ij}=g\frac{2e^2}{Mh}$, as well as the the power-law of the temperature dependent corrections $\sim{T}^{4g(M-1)/{M}-2}$.
For stronger repulsive interactions $\frac{M-1}{2M}<g<\frac{M}{2(M-1)}$ a stable fixed point occurs at intermediate coupling strengths (see Fig.~5).

To conclude, our work predicts a set of remarkable transport phenomena in multi-terminal Majorana islands in the vicinity of the charge-degeneracy point, including a universal fractional quantum conductance at zero temperature, and its universal power-law correction at low temperature. Observation of such phenomena will clearly demonstrate the Majorana nature of zero modes in a superconductor island, defined by the operator algebra Eq.~\eqref{gamma} and acting as a charge-neutral Fermi-Bose transformer.

%###########################
%###########################

{\it Acknowledgements:} We thank Ian Affleck, Moshe Goldstein, Charlie Kane, Charlie Marcus and Michal Papaj for helpful discussions. This work is supported by David and Lucile Packard Foundation (LF), Israel Science Foundation Grant No.~1243/13, and the Marie Curie CIG Grant No.~618188 (ES), as well as the the Minerva Foundation (KM).

\appendix

\section{Detailed study of the phase diagram}

In the main text we described the mapping of Majorana islands coupled to $M$ leads onto a QBM model. Here we elaborate on the various  steps of the derivation and the analysis of the phase digram (Fig.~5).     We start with bosonization of the leads and integration of all degrees of freedom away from $x=0$. The resulting effective action describes a particle subject to a periodic potential, and coupled to a dissipative bath. Within this QBM model we calculate the scaling dimensions of various allowed perturbations, and study the weak- and strong-tunneling limits  near the charge-degeneracy point and in the Kondo regime.

\subsection{Bosonization}

We start the bosonization procedure by mapping the model system described above onto a spin chain. For this purpose, we describe the leads as chains of fermions
\begin{equation}\label{Appendix-2}
H_{\rm{lead}}= \mathcal{J}\sum_{j=1}^M \sum_{m=1}^{\infty} c^\dagger_{j,m} c_{j-1,m}+{\rm{H.c.}},
\end{equation}
where the lattice constant is set to unity and the hopping parameter $\mathcal{J}$ is fixed to reproduce the density of states in the leads $\rho=(4 \pi \mathcal{J})^{-1}$. The creation (annihilation) operator at the boundary site $m=0$ is identified with the boundary field operator $\psi_j^{\dag} (0)= c_{j,0}^{\dag}$ ($\psi_j(0)= c_{j,0}$). In general, the standard Jordan-Wigner transformation that maps one-dimensional fermions onto a spin chain fails for a system of $M>2$ semi-infinite wires joined at a single point. When the one-dimensional wires are coupled to the Majorana island, we can define commuting spin operators as a product of electron operators in the leads and the corresponding Majorana mode operators:
\begin{equation}\label{Appendix-1}
S^+_{j,m} = c^\dagger_{j,m} \gamma_j.
\end{equation}
Correspondingly, the Hamilton can be expressed in terms of $M$ $xy$-spin chains, all connected at the origin to the spin operator of the island:
\begin{equation}\label{Appendix0}
H =\sum_{j=1}^M\left[\mathcal{J}  \sum_{m=1}^{\infty} S^+_{j,m} S^-_{j,m+1}+t_j S^+_{j,1} \sigma^-+{\rm{H.c.}}\right]+{\Delta_g} \sigma^z.
\end{equation}
In this description of the system, the Majorana operators disappear from the Hamiltonian, manifesting the Bose statistics of the charge carriers in the leads.

Next we express the spin operators in each chain in terms of  left ($\varphi^L$) and right ($\varphi^R$) moving chiral modes $S_j^{+}(x)\sim e^{i k_F x} e^{i \varphi_j^{R}(x)}+e^{-i k_F x} e^{i \varphi_j^{L}(x)}$, where $0\leq{x}<\infty$. However, we find it more convenient to describe each lead as an infinite chain $-\infty<x<\infty$, and express the spin operators in term of a \textit{single}  chiral mode:
\be\label{Appendix1}
S_j^{+}(x)\sim e^{i k_F x} e^{i \varphi_j(x)},
\ee
where $\varphi_j(x)=\varphi_j^L(x) \theta(x) + \varphi_j^R(-x) \theta(-x)$. The chiral operators obey the commutation relations $[\varphi_i(x) , \partial_x \varphi_j(x')] = 2 \pi i \delta(x-x')\delta_{ij}$, and the conjugate operators can be identified  as the electron density operators $ \rho_j(x) =\frac{1}{2\pi}\partial_x \varphi_j(x)$. This is because $e^{i\varphi_j(x)}$ changes the total charge by $1$, and  similarly  $e^{2\pi {i}\rho_j(x)}$ shifts the phase by $2\pi$.  The imaginary-time action of the leads,  corresponding to the first term in  Eq.~\eqref{Appendix0}, can be written in terms of the phase fields $\varphi_j(x,\tau)$ as
\begin{align} \label{Appendix2}
&S_{\text{leads}} =\\\nonumber
& -\frac{1}{4 \pi}\sum_{j=1}^{M}\int_{0}^{\beta}\hspace{-1mm}d\tau\hspace{-1mm}\int dx\left[  \varphi_j(x,\tau) \partial_x (v\partial_x -i\partial_{\tau}) \varphi_j(x,\tau)\right].
\end{align}
Here, $\beta=T^{-1}$ is the inverse temperature, and $v$ is the Fermi velocity.

The scaling dimension of the spin operators $\Delta_{s}$ is obtained from the  zero-temperature correlation function of the field $\varphi_j(x,\tau)$ as
\begin{align}\label{Appendix3}
&\langle{S}_j^{-}(x,\tau)S_j^{+}(0,0)\rangle=\langle{e}^{-i\varphi_j(x,\tau)}{e}^{i\varphi_j(0,0)}\rangle\\\nonumber&= e^{-\frac{1}{2}\langle\left[\varphi_j(x,\tau)-\varphi_j(0,0)\right]^2\rangle}\propto\left({x-iv\tau}\right)^{-2\Delta_{s}}.
\end{align}
From the action Eq.~\eqref{Appendix2}, we get that
\begin{align}\label{Appendix4}
\langle\left[\varphi_j(x,\tau)-\varphi_j(0,0)\right]^2\rangle=2\log\left(x-i\tau\right)+\text{const},
\end{align}
and  $\Delta_{s}=\frac{1}{2}$.

Up to here, we considered free electrons in the leads. To generalize the derivation to interacting leads, we introduce the Luttinger parameter $g$ into the action:
\begin{align} \label{Appendix5}
&S_{\text{leads}} =\\\nonumber
& -\frac{g}{4 \pi}\sum_{j=1}^{M}\int_{0}^{\beta}\hspace{-1mm}d\tau\hspace{-1mm}\int dx\left[  \varphi_j(x,\tau) \partial_x (v\partial_x -i\partial_{\tau}) \varphi_j(x,\tau)\right].
\end{align}

Here $g<1$ ($g>1$) corresponds to repulsive (attractive) interactions. Consequently, the zero-temperature correlation function
\begin{align} \label{Appendix6}
\langle\left[\varphi_j(x,\tau)-\varphi_j(0,0)\right]^2\rangle=\frac{2}{g}\log\left(x-i\tau\right)+\text{const},
\end{align}
and the scaling dimension of the spin operators is $\Delta_{s}=\frac{1}{2g}$. Furthermore, the definition of the conjugate fields is also $g$-dependent, $[\varphi_j(x'),\partial_{x}\varphi_i(x)]=2\pi {g}^{-1}{i}\delta(x-x')\delta_{ij}$, and the density operator  becomes $ \rho_j(x) =\frac{g}{2\pi}\partial_x \varphi_j(x)$.

In the derivation of the action given by Eq.~\eqref{Appendix5} as well as of the properties of $\varphi(x,\tau)$ we followed  Ref.~\cite{vonDelft}. An alternative approach would be to perform the bosonization with the  non-chiral operators $\phi$ and $\theta$ (see for example Ref.~\cite{Giamarchi}) and use the relations:
\begin{align} \label{Appendix7}
&\phi_j(x) = \frac{\varphi_j^{R}(x)+\varphi_j^{L}(x)}{\sqrt{2}}; \\\nonumber
&\theta_j(x) = \frac{\varphi_j^{R}(x)-\varphi_j^{L}(x)}{\sqrt{2}}.
\end{align}
Here $\varphi_j^{L}$ ($\varphi_j^{R}$) is the left (right) chiral operator. The left- and right-chiral fields are connected through the transformation  $x \to -x$.

finally, we turn to the coupling term between the leads and the Majorana island, second term in Eq.~\eqref{Appendix0}. Using the expressions for the spin lowering and raising operators  in terms of the chiral fields,  the tunneling Hamiltonian becomes
\begin{align} \label{Appendix8}
H_{\rm{T}} &=\sum_{j=1}^M t_je^{i\varphi_j(0)}  \sigma^{-}+{\rm{H.c.}}.
\end{align}

\subsection{Boundary Action}

The next step in the mapping onto QBM is to integrate out the degrees of freedom away from $x=0$. For this purpose, we use the Fourier decomposition of the fields:
\be \label{Appendix9}
\varphi_j(x,\tau)=\beta^{-1}\sum_{n}\int\frac{dk}{2\pi} e^{i (k x + \omega_n \tau)}\varphi_j(k,\omega_n),
\ee
and the corresponding action:
\be \label{Appendix10}
S=\frac{g}{4 \pi \beta} \sum_{j,n}\int\frac{dk}{2\pi}k\left(vk-i\omega_n\right) |\varphi_j(k,\omega_n)|^2.
\ee
Here, $\omega_n=2\pi{n}\beta^{-1}$ are the Matsubara frequencies. The field at the boundary ($x=0$) is obtained by integrating over momentum
\be \label{Appendix11}
\varphi_j(\omega_n)=\int\frac{dk}{2\pi} \varphi_j(k,\omega_n).
\ee
To find the  boundary action $S= \frac{1}{2}\sum_{j,n} G_{jj}^{-1}(\omega_n)$ $\times|\varphi_j(\omega_n)|^2$, we have to calculate the correlation function $G_{jj}=\langle\varphi_j(\omega_n)\varphi_j(-\omega_n)\rangle$:
\be \label{Appendix12}
G_{jj}(\omega_n)=\frac{ \beta}{g}\int\frac{dk}{k\left(vk-i\omega_n\right) }=\frac{\pi\beta}{g|\omega_n|}.
\ee
The action for the boundary field:
\be\label{Appendix13}
S=\frac{g}{ 2\pi\beta}\sum_{n,j}{|\omega_n| }|\varphi_j(\omega_n)|^2
\ee
coincides with the expression given in Eq.~\eqref{Dissipation} when $g=1$ and $\beta\rightarrow\infty$. Eq.~\eqref{Appendix13} describes a particle  subject to a classical friction term~\cite{CL} (Ohmic dissipation). Thus, the particle exhibits Brownian motion in an $M$-dimensional space, where the field $\varphi_j$ is its momentum  along the $j$-axis.

In the QBM framework, tunneling between the leads and the Majorana island is encoded in terms of the form $e^{i\sqrt{2}\vec{\varphi}(0)\cdot\vec{R}}$. Therefore, to analyze the phase diagram it is important to find the scaling dimension of such terms. Setting $x=0$ in Eq.~\eqref{Appendix3}, we find the scaling dimensions of  $e^{i\sqrt{2}\vec{\varphi}(0)\cdot\vec{R}}$ to be
\begin{align}\label{Appendix14}
\Delta\left[e^{i\sqrt{2}\vec{\varphi}(0)\cdot\vec{R}}\right]=\frac{|\vec{R}|^2}{g}.
\end{align}
This expression is needed for the analysis of the RG flow for the QBM in the weak-tunneling (strong periodic potential) regime.

For the strong-tunneling limit, we need to find the scaling dimensions of terms of the form $e^{i\vec{r}\cdot\vec{G}}$ that shift $\sqrt{2}\varphi_j(0)$ by $2\pi G_j$. Previously, we saw that a shift of the phase in an infinite lead  is generated by the density operator $\rho_j(x)$. On the boundary, we note that the operator $r_j=\sqrt{2}\pi\lim_{\epsilon\rightarrow0}\int_{-\epsilon}^{\epsilon}dx\rho_j(x)$ satisfies the desired commutation relations:
\begin{align}\label{Appendix15}
[\sqrt{2}\varphi_i(0),r_j]=2\pi\lim_{\epsilon\rightarrow0}\int_{-\epsilon}^{\epsilon}dx[\varphi_i(0),\rho_j(x)]=2\pi{i}\delta_{i,j}.
\end{align}
The definition of the density operator, allows us to rewrite this operator in terms of the phase $\varphi(x)$ as   $r_j=\frac{g}{\sqrt{2}}\lim_{\epsilon\rightarrow0}\left[\varphi_j(\epsilon)-\varphi_j(-\epsilon)\right]$. We note that in the strong-tunneling limit, $\varphi_j(0)$ has a finite expectation value, and correspondingly $\lim_{\epsilon\rightarrow0}\left[\varphi_j(\epsilon)+\varphi_j(-\epsilon)\right]=\text{const}$ (for formulation of the boundary condition in terms of the non-chiral operators see Ref.~\cite{Chamon}). This property reflects the fact that each electron that comes from $x=\infty$ is transferred to the Majorana island. As a result,
 \begin{align}\label{Appendix16}
r_j=\sqrt{2}g\lim_{\epsilon\rightarrow0}\varphi_j(\epsilon)=\sqrt{2}g\varphi_j(0)+\text{const},
\end{align}
we find that scaling dimension of $e^{i\vec{r}\cdot\vec{G}}$ is
 \begin{align}\label{Appendix17}
\Delta\left[e^{i\vec{r}\cdot\vec{G}}\right]=g|\vec{G}|^2.
\end{align}

\subsection{RG flow diagram near the charge-degeneracy point}

The bosonized description of a Majorana island coupled to $M$ leads given in Eq.~\eqref{Appendix5} and the corresponding boundary action in Eq.~\eqref{Appendix13} allow us to analyze the phase diagram of the system. To follow the derivation in the main text, we assume equal coupling constants to all leads and that the gate voltage is tuned to the charge-degeneracy point, $\Delta_g=0$. The starting point of the calculation is the full Hamiltonian before integrating out fluctuations away from $x=0$:
\begin{align}\nonumber
H&= \sum_{j=1}^{M}\int{dx}\left\{\frac{vg}{4 \pi}(\partial_x \varphi_j(x))^2+\frac{v}{2}J_z \sigma^z \partial_x \varphi_j(x) \delta(x)\right.\\
&\left.+ J_{\perp}  \left(e^{i \varphi_j(x)}\sigma^- + {\rm{H.c.}}\right) \delta(x)
\phantom{\frac{vg}{4 \pi}}\hspace{-5mm}\right\}.\label{Appendix18}
\end{align}
Although the bare Hamiltonian does not include the $J_z$ term, such a term is generated in the RG process. In the previous sections we showed that the scaling dimension of the (bare) tunneling operator is $\frac{1}{2g}$, however, it is expected to change in the RG process. To find the renormalized scaling dimension of the tunneling term, we rewrite the above Hamiltonian in the following form:
\begin{align}\nonumber
H&= \sum_{j=1}^{M}\int{dx}\left\{\frac{vg}{4 \pi}\left(\partial_x \varphi_j(x)+\pi{g}^{-1}{J}_z \sigma^z \delta(x)\right)^2\right.\\
&\left.+ J_{\perp} \left(e^{i \varphi_j(x)}\sigma^- + {\rm{H.c.}}\right) \delta(x)
\phantom{\frac{v}{4 \pi}}\hspace{-5mm}\right\}.\label{Appendix19}
\end{align}
The $J_z$-term can be eliminated from the Hamiltonian by the unitary transformation:
\begin{align}\label{Appendix20}
U=e^{\frac{i}{2}J_z\sigma_z\sum_{j}\varphi_j}.
\end{align}
Under this transformation $U^{\dag}\sigma^{-}U=\sigma^{-}\exp[iJ_z\sum_{j}\varphi_j]$, and the Hamiltonian becomes:
\begin{align}\label{Appendix21}
\tilde{H}&=U^{\dag}HU= \sum_{j=1}^{M}\int{dx}\left\{\frac{vg}{4 \pi}(\partial_x \varphi_j(x))^2\right.\\\nonumber
&\left.
+ J_{\perp}  \left(e^{i \varphi_j(x)-iJ_z\sum_{\ell}\varphi_{\ell}(x)}\sigma^- + {\rm{H.c.}}\right) \delta(x)\right\}.
\end{align}
Specifically, for $J_z=1/M$ the center-of-mass boson field $\sum_j\varphi_j$ drops out from the tunneling term.  Since the flow of $J_z$ stops at $1/M$, the system reaches a new (Toulouse-like) fixed point.

At the fixed point, we integrate out the degrees of freedom away from $x=0$, and write the boundary action as:
\begin{align}\label{Appendix22}
\tilde{S} =\beta^{-1}\sum_{n,j} \hspace{-0.5mm}\left\{ \hspace{-0.5mm}\frac{g|\omega_n|}{2\pi}|\varphi_j(\omega_n)|^2 \hspace{-1mm}+\hspace{-0.5mm}J_{\perp}\hspace{-1mm}\left(e^{i\sqrt{2}\vec{ \varphi_j}(\omega_n)\cdot\vec{R}_{\parallel}^{(j)}}\hspace{-2mm}+ {\rm{H.c.}}\hspace{-0.5mm}\right)\hspace{-1mm}\right\}\hspace{-0.5mm},
\end{align}
where $(R_{\parallel}^{j})_{i}=\frac{1}{\sqrt{2}}\left[\delta_{ij}-\frac{1}{M}\right]$ is the $i$-th component of the vector $\vec{R}_{\parallel}^{j}$. Thus, in the Toulouse-like fixed point the action describes  QBM of a particle that is subject to a periodic potential in an $M-1$  dimensional space spanned by $\vec{R}_{\parallel}^{j}$.  From Eq.~\eqref{Appendix14} we find that the scaling dimension of the tunneling term is given by
\begin{align}\label{Appendix23}
\Delta\left[e^{i\sqrt{2}\vec{ \varphi_j}(0)\cdot\vec{R}_{\parallel}^{(j)}}\right]=\frac{1}{2g}\left(1-\frac{1}{M}\right).
\end{align}
For free electrons in the leads ($g=1$), the scaling dimension of the tunneling term is relevant. Therefore, the weak-tunneling regime is unstable, and $J_{\perp}$ flows to infinity.  At this fixed point the lattice potential vanishes, and the particle can move freely, i.e.,  charge strongly fluctuates between the leads. Correspondingly, the potential for $\vec{\varphi}$ is maximal and the field is locked to one of its minima.

To analyze the stability of this new fixed point we note that the symmetry allowed perturbations are of the form
\begin{align}\label{Appendix24}
e^{i\vec{r}\cdot\vec{G}},
\end{align}
where $\vec{G}$ is a reciprocal vector of the lattice spanned by $\vec{R}_{\parallel}$. This kind of terms restore the lattice potential that vanished in the RG flow. Equivalently, such terms describe tunneling between minima of the potential for $\vec{\varphi}$, and they tend to decouple the leads from the Majorana island, i.e., to pin the charge. The scaling dimension of the perturbation in Eq.~\eqref{Appendix24} was calculated in the previous section (see Eq.~\eqref{Appendix18}). To find the reciprocal lattice vector, we use the relation $\vec{G}\cdot\vec{\mathcal{R}}=\text{integer}$ for any Bravais lattice vector $\vec{\mathcal{R}}$. For the $M-1$  dimensional lattice defined by $\vec{R}_{\parallel}$, the Bravais  vectors are  $\left(\mathcal{R}^{(i,j)}\right)_{\ell}=\frac{1}{\sqrt{2}}\left(\delta_{i,\ell}-\delta_{j,\ell}\right)$. Correspondingly, the shortest reciprocal lattice vectors are:
\begin{align}\label{Appendix25}
\left(G^{(j)}\right)_{i}=\sqrt{2}\left(\delta_{i,j}-\frac{1}{M}\right),
\end{align}
and the scaling dimension of the tunneling operator on the reciprocal lattice is
\begin{align}\label{Appendix26}
\Delta\left[e^{i\vec{r}\cdot\vec{G}}\right]=2g\left(1-\frac{1}{M}\right).
\end{align}
Therefore, the leading perturbation is irrelevant for free leads.

For interacting leads, the tunneling term in Eq.~\eqref{Appendix23} is relevant  for $g>\frac{M-1}{2M}$, and the periodic potential term in Eq.~\eqref{Appendix26} is relevant for  $g<\frac{M}{2(M-1)}$. Therefore, for  $g>\frac{M}{2(M-1)}$ the system flows to the strong-tunneling (vanishing potential) fixed point, while the strong potential (decoupled leads and island) fixed point is stable for $g<\frac{M-1}{2M}$. As shown in Fig.~5, a stable intermediate fixed point appears for $\frac{M-1}{2M}<g<\frac{M}{2(M-1)}$.

\subsection{RG flow diagram in the Kondo regime}

When the gate voltage is tuned far from $\Delta_g=0$, charge fluctuations in the island are gapped.  As a result, electrons can only hop between the leads via virtual transitions through the island, and the effective action becomes:
\begin{align}\label{Appendix27}
H&=\int{dx}\left\{\frac{vg}{4 \pi} \sum_{j=1}^{M}(\partial_x \varphi_j(x))^2\right.\\\nonumber
&\left.\hspace{5mm}+\lambda \sum_{i\neq{j}} \left(e^{i( \varphi_i(x)- \varphi_j(x))} + {\rm{H.c.}}\right) \delta(x)\right\},
\end{align}
Where $\lambda=\frac{|t|^2}{\Delta_g}$. Here, no new terms are generated in the RG process, and the center-of-mass boson $\sum_{j}\varphi_j$ does not appear in the tunneling term.  Therefore to obtain the flow diagram in the Kondo regime, we follow the steps introduced in the previous section after eliminating the $J_z$ term (starting at Eq~\eqref{Appendix22}). The boundary action can be written as
\begin{align}\label{Appendix28}
H_{K}&=\lambda \sum_{i\neq{j}} \left(e^{i\sqrt{2}\vec{R}^{ij}\cdot\vec{\varphi}(0)} + {\rm{H.c.}}\right),
\end{align}
where $\left(R^{ij}\right)_{\ell}=\frac{1}{\sqrt{2}}\left(\delta_{i,\ell}-\delta_{j,\ell}\right)$. From Eq.~\eqref{Appendix14}, we find that the scaling dimension of the tunneling operator is $\frac{1}{g}$. As a result, the tunneling term is marginally relevant for free leads and the system flows to the weak-potential limit. In the presence of arbitrarily weak repulsive interactions, the tunneling term is irrelevant~\cite{AltlandEgger} and, for not too strong bare tunnel couplings, the leads decouple from the island.

Interestingly, $\vec{R}^{ij}$ in the Kondo limit are the Bravais vectors of the lattice near the charge-degeneracy point (see discussion below Eq.~\eqref{Appendix24}). Near the charge-degeneracy point, however, the lattice is defined by a basis vector in additional to the Bravais vectors. This point is illustrated in Fig.~4 where the QBM near $\Delta_g=0$ is on a honeycomb lattice, while in the Kondo regime, the QBM is confined to a plane of constant total charge, and the periodic potential is triangular. As a result, the reciprocal lattice vectors in both cases are identical, and so is the scaling dimension of the leading operator in the strong-tunneling limit, Eq.~\eqref{Appendix26}. We conclude that for noninteracting leads, the strong-tunneling fixed points are the same in the vicinity and far from $\Delta_g=0$. However, only near the charge-degeneracy point  the fixed point remains stable in the presence of weak repulsive interactions.

\bibliographystyle{apsrev}

\begin{thebibliography}{10}
	
\bibitem{kitaev} A. Kitaev, \textit{Ann. Phys. (N.Y.)} {\bf 303}, 2 (2003).
	
\bibitem{read} N. Read and D. Green, \textit{Phys. Rev. B.} {\bf 61}, 10267 (2000).

\bibitem{wilczek} F. Wilczek, \textit{Nat. Phys.} {\bf 5}, 614 (2009).

\bibitem{fu} 	L. Fu, \textit{Phys. Rev. Lett.} {\bf 104}, 056402 (2010).
	
\bibitem{xu} C. Xu, and L. Fu, \textit{Phys. Rev. B.} {\bf 81}, 134435 (2010).

\bibitem{hassler} F. Hassler, A. R. Akhmerov, C. W. J. Beenakker, \textit{New J. Phys.} {\bf 13}, 095004 (2011).
	
\bibitem{heck} B. van Heck, A. R. Akhmerov, F. Hassler, M.Burrello, and C. W. J. Beenakker, \textit{New J. Phys.} {\bf 14}, 035019 (2012).
				
\bibitem{terhal}	B. M. Terhal, F. Hassler, and D. P. DiVincenzo, \textit{Phys. Rev. Lett.} {\bf 108}, 260504 (2012).

\bibitem{hyart} T. Hyart, B. van Heck, I. C. Fulga, M. Burrello, A. R. Akhmerov, and C. W. J. Beenakker, \textit{Phys. Rev. B.} {\bf 88}, 035121 (2013).

\bibitem{grosfeld} E. Ginossar, E. Grosfeld, \textit{Nat. Commun.}  {\bf 5}, 4772 (2014). 	

\bibitem{vijay}	S. Vijay, T. H. Hsieh, and L. Fu,  \textit{Phys. Rev. X.} {\bf 5}, 041038 (2015).

\bibitem{alicea} D. Aasen {\it et al},  arXiv:1511.05153
	
\bibitem{landau} L. A. Landau, S. Plugge, E. Sela, A. Altland, S. M. Albrecht, and R. Egger, \textit{Phys. Rev. Lett.} {\bf 116}, 050501 (2016).
	
\bibitem{egger}	A. Zazunov, A. Levy Yeyati, and R. Egger, \textit{Phys. Rev. B.} {\bf 84}, 165440 (2011).

\bibitem{glazman} B. van Heck, R.M. Lutchyn, and L.I. Glazman, arXiv:1603.08258

\bibitem{marcus} S. M. Albrecht, A. P. Higginbotham, M. Madsen, F. Kuemmeth, T. S. Jespersen, J. Nyg{\aa}rd, P. Krogstrup, and C. M. Marcus, \textit{Nature} {\bf 531}, 206 (2016).
	

\bibitem{sarma} R. M. Lutchyn, J. D. Sau, and S. Das Sarma,  \textit{Phys. Rev. Lett.} {\bf 105}, 077001 (2010).
	
\bibitem{oreg} Y. Oreg, G. Refael, and F. von Oppen,  \textit{Phys. Rev. Lett.} {\bf 105}, 177002 (2010).

\bibitem{kouwenhoven} V. Mourik, K. Zuo, S. M. Frolov, S. R. Plissard, E. P. A. M. Bakkers, and L. P. Kouwenhoven,  \textit{Science} {\bf 336}, 1003 (2012).

\bibitem{weizmann} A. Das, Y. Ronen, Y. Most, Y. Oreg, M. Heiblum, and H. Shtrikman,  \textit{Nat. Phys.} {\bf 8}, 887 (2012).
	

	
	

\bibitem{marcus2} P. Krogstrup {\it et al}, \textit{Nat. Mat.} {\bf 14}, 400  (2015).


\bibitem{marcus3} C. M. Marcus, private communication


\bibitem{BeriCooper} B. B{\'e}ri, and N. Cooper, \textit{Phys. Rev. Lett.} {\bf 109}, 156803 (2012).

\bibitem{AltlandEgger} A. Altland, and R. Egger, \textit{Phys. Rev. Lett.} {\bf 110}, 196401 (2013).

\bibitem{Beri} B. B{\'e}ri, \textit{Phys. Rev. Lett.} {\bf 110}, 216803 (2013).
	
\bibitem{Affleck13} I. Affleck, and D. Giuliano, \textit{J. Stat. Mech.} {\bf 2013} P06011 (2013).

\bibitem{Tsvelik} A. Altland, B. B{\'e}ri, R. Egger, and A. M. Tsvelik, \textit{J. Phys. A.} {\bf 47}, 265001 (2014).

\bibitem{AltlandBeryEggerTsvelik14} A. Altland, B. B{\'e}ri, R. Egger, and A. M. Tsvelik, \textit{Phys. Rev. Lett.} {\bf 113}, 076401 (2014).

\bibitem{numerics} M. R. Galpin, A. K. Mitchell, J. Temaismithi, D. E. Logan, B. B{\'e}ri, N. R. Cooper, \textit{Phys. Rev. B.} {\bf 89}, 045143 (2014).

\bibitem{ZazunovAltlandEgger14} A. Zazunov, A. Altland, and R. Egger, \textit{New J. Phys.} {\bf 16}, 015010 (2014).

\bibitem{Eriksson14a} E. Eriksson, A. Nava, C. Mora, and R. Egger, \textit{Phys. Rev. B.} {\bf 90}, 245417  (2014).

\bibitem{Eriksson14}	E. Eriksson, C. Mora, A. Zazunov, and R. Egger, \textit{Phys. Rev. Lett.} {\bf 113}, 076404 (2014).

\bibitem{Kashuba15}  O. Kashuba and C. Timm, 	\textit{Phys. Rev. Lett.} {\bf 114}, 116801 (2015).

\bibitem{Pikulin16} D. I. Pikulin, Y. Komijani, and I. Affleck, \textit{Phys. Rev. B.} {\bf 93}, 205430 (2016).

\bibitem{Meidan16} D. Meidan, A. Romito, and P. W. Brouwer, \textit{Phys. Rev. B.} 	{\bf 93}, 125433 (2016).

\bibitem{Plugge16} S. Plugge, A. Zazunov, E. Eriksson, A. M. Tsvelik, and R. Egger, \textit{Phys. Rev. B.} {\bf 93}, 104524 (2016).

\bibitem{YiKane} H. Yi, and C. L. Kane, \textit{Phys. Rev. B.} {\bf 57} R5579 (1998).

\bibitem{Chamon} C. Chamon, M. Oshikawa, and I. Affleck, \textit{Phys. Rev. Lett.} {\bf 91} 206403 (2003); \textit{J. Stat. Mech.} {\bf 602} P02008 (2006).
	
\bibitem{Nayak} C. Nayak, M. P. A. Fisher, A. W. W. Ludwig, and H. H. Lin,  \textit{Phys. Rev. B.} {\bf 59} 15694 (1999).
	
\bibitem{KaneFisher} C.~Kane, and M.~P.~A.~Fisher,  \textit{Phys. Rev. B.} {\bf 46}, 15233 (1992).

\bibitem{vonDelft} J. von Delft, and H. Schoeller,  \textit{Annal. Phys.} {\bf 7} 225 (1998).	
	
\bibitem{Giamarchi} T. Giamarchi, \textit{Quantum physics in one dimension}    (Carendon, Oxford, 2003).
	
\bibitem{CL}	A.~O.~Caldeira, and A.~J.~Leggett,   \textit{Phys. Rev. Lett.}~{\bf46}, 211 (1981).



	
\end{thebibliography}

\end{document}